\documentstyle[prl,multicol,aps,epsfig]{revtex}
\begin{document}

\title{Chaotic Scattering on Graphs}
\author{Tsampikos Kottos$^1$ and Uzy Smilansky$^2$ \\
$^1$ Max-Planck-Institut f\"ur Str\"omungsforschung, 37073 G\"ottingen,
Germany,\\
$^2$ Department of  Physics of Complex Systems,
The Weizmann Institute of Science, 76100 Rehovot, Israel}
\date{\today }
\maketitle
\bigskip

\begin{abstract}
Quantized, compact graphs were shown to be excellent paradigms for quantum
chaos in bounded systems. Connecting them with leads to infinity we show that
they display  all the features which characterize  scattering systems with an
underlying classical chaotic dynamics.  We derive exact expressions for the
scattering matrix, and an exact trace formula for the density of resonances,
in terms of classical orbits, analogous to the semiclassical theory of chaotic
scattering. A statistical analysis of the cross sections and resonance
parameters compares well with the predictions of Random Matrix Theory. Hence,
this system is proposed as a convenient tool to study the generic behavior of
chaotic scattering systems, and their semiclassical description.
\\
\end{abstract}

\hspace {1.5 cm} PACS numbers: 05.45.+b, 03.65.Sq

\begin{multicols}{2}
Quantum graphs  provide a very useful tool to study bounded quantum systems
which are chaotic in the classical limit \cite {KS97}. Here, by attaching
infinite leads we turn the compact graphs into scattering systems, and show
that they display chaotic scattering \cite {S89,molecular}, a phenomenon with
applications in many fields, ranging from nuclear \cite {nuclear}, atomic
\cite {atomic} and molecular \cite {molecular} physics, to mesoscopics
\cite {mesoscopics} and classical wave scattering \cite{electro}. We express
the {\it quantum} scattering matrix and the trace formula for the density
of its resonances in terms of the orbits of the underlying {\it classical}
scattering system. These expressions are the exact analogues of the
corresponding semiclassical approximations available in the theory of chaotic
scattering \cite {S89,GR89,M73}. With these tools we analyze the distribution
of resonances and the statistics of the fluctuating scattering amplitudes and
cross sections. We show that graphs provide new insight on the connection
between Random Matrix Theory (RMT) and chaotic scattering. Moreover, we
illustrate their advantages as versatile and convenient tools for numerical
studies of chaotic scattering.

Consider first a {\it bounded} graph $ {\cal G}$. It consists of $V$ {\it
vertices} connected by {\it bonds}. The valency $v_i$ of a vertex $i$
is the number of bonds which emanate from it, and we allow only a single
bond between any two vertices. The total number of bonds is $B={1\over 2}
\sum_{i=1} ^V v_{i}$. The bond connecting the vertices $i$ and $j$ is
denoted by $b \equiv (i,j)$.  We shall also distinguish between the directions
on the bond using  $\hat b \equiv (j,i)$ to denote the reverse direction.
The length of the bonds are denoted by $L_{b}$ and we shall henceforth assume
that they are {\it rationally independent}. In the {\it directed-bond}
notation  $L_{b} =L_{\hat b}$. The {\it scattering} graph ${\tilde {\cal G}}$
is obtained by adding infinite leads at the vertices of ${\cal G}$, changing
their valency to ${\tilde v}_i=v_i+1$. The leads are distinguished by the
index $i$ of the vertex to which they are attached.

The Schr\"odinger operator on the graph consists of the one dimensional
Laplacian
$\left(- i {\rm d}_x - A_b  \right ) ^2$ supplemented by boundary conditions on
the vertices \cite {KS97}. The ``vector potentials" $A_{b}=-A_{\hat b}$  are
introduced to break time reversal symmetry, and their value may vary from
bond to
bond. On each of the bonds $b$ or leads $i$  the wave function is expressed in
terms of counter propagating waves with a wave number $k$:
\begin {eqnarray}
&&{\rm On\ the \  bonds:\ } \psi_{b} = a_{b} {\rm e}^{i(k+A_{b})x_{b}}
+a_{\hat b} {\rm e}^{i(k+A_{b})(L_b-x_{b})} \nonumber  \\
&&{\rm On\  the \ leads\ :\ } \psi_i =  I_{i} {\rm e}^{-ikx_{i}} + O _{i} {\rm
e}^{ ikx_{i}}
\end{eqnarray}
where the coordinate $x_b$ on the bond $b=(i,j)$ takes the value $0$ ($L_b$)
at the vertex $i$ ($j$) while $x_i$ measures the distance from the vertex
along the
lead $i$.

The  amplitudes  $a_b,a_{\hat b}$ on the bonds and $I_i,O_i$ on the leads are
determined by matching conditions  at the vertices. They are expressed in
terms
of the  $\tilde v_i \times \tilde v_i $ {\it  vertex scattering matrices}
$\Sigma^{(i)}_{j,j'}$, where $j,j'$ go over all the $v_i$ bonds and the lead
which emanate from $i$. The $\Sigma^{(i)}$ are symmetric {\it unitary}
matrices,
which guarantee current conservation at each vertex by requiring
\begin{eqnarray}
{\left (
\begin {array} {c}
  O_i  \\  a_{i,j_1}\\ \cdot  \\   a_{i,j_{v_i}}
 \end {array}
\right )} =
{\left (
\begin{array}{cccc}
\rho^{(i)}  &  \tau^{(i)}_{j_1} & \cdot   & \tau^{(i)}_{j_{v_i}} \\
\tau^{(i)}_{j_1} & \tilde \sigma^{(i)}_{j_1,j_1} & \cdot  &
 \tilde \sigma^{(i)}_{j_1,j_{v_i}} \\
\cdot  &      \cdot &\cdot &\cdot \\
\tau^{(i)}_{j_{v_i}} & \tilde \sigma^{(i)}_{{j_{v_i}},j_1} &
  \cdot & \tilde \sigma^{(i)}_{j_{v_i},j_{v_i}} \\
\end{array}
\right )}
{\left ( \begin{array} {c}
  I_i  \\ c_{j_1,i} \\ \cdot  \\ c_{j_{v_i},i}
 \end {array} \right ) } \ \
\label{vertexcond}
\end{eqnarray}
where $c_{j_l,i}=a_{j_l,i}e^{i (k+A_{(i,j_l)})L_{(i,j_l)} }$. Above, the vertex
scattering matrix $\Sigma^{(i)}$ was written explicitly in terms of the vertex
reflection amplitude  $\rho^{(i)}$, the lead-bond transmission amplitudes
$\{ \tau^{(i)}_{j}\}$, and the $v_i\times v_i$   bond-bond transition
matrix $\tilde \sigma^{(i)}_{j,j'}$, which is {\it sub unitary} ($|\det
\tilde
\sigma^{(i)}|<1 $), due to the coupling to the leads.  As an example, for
$v$-regular graphs
($v_i=v\forall i$) with Neumann matching conditions on the vertices, \cite
{KS97}:
\begin {eqnarray}
\tilde \sigma^{(i)} _{j,j'}= {2 \over v+1}  -\delta _{j,j'}\ ;
\tau^{(i)}_j =   {2 \over v+1};\  \rho^{(i)} ={2 \over v+1}-1 \ .
\label{Neumann}
\end{eqnarray}
Combining the equations (\ref {vertexcond}) for all the vertices, we obtain the
$V\times V$ scattering matrix $S^{(V)}$ which relates the outgoing and
incoming
amplitudes on the leads,
\begin{eqnarray}
\label{scatmat}
S^{(V)}_{i,j} & =& \delta_{i,j} \rho^{(i)}  \\
 &+& \sum_{r,s} \tau^{(i)}_r
\left (I-\tilde S(k;A) \right )^{-1}_{(i,r),(s,j)} D_{(s,j)}
\tau _s^{(j)}. \nonumber
\end{eqnarray}
Here,  $D_{(s,j)}(k;A) = \exp ( i(k+A_{(s,j)})L_{(s,j)} ) $ is a diagonal
matrix in
the $2B$ space of directed bonds. The matrix  $\tilde S(k;A)= D(k;A) \tilde R$
propagates the wave functions:  $\tilde R$  assigns a scattering amplitude for
transitions between connected directed bonds: $\tilde R_{(i,r),(s,j)} =
\delta_{r,s} \tilde \sigma ^{(r)}_{i,j}$;  $D(k;A)$ provides the phase due to
free propagation. The matrix $\tilde R$ is sub-unitary, since $|\det \tilde R|
= \prod_{i=1}^V |\det \tilde \sigma ^{(i)}|<1$. The scattering matrix
(\ref{scatmat}) is interpreted in the following way. The prompt reflection at
the entrance vertex induces a ``direct" component. The ``chaotic" component
starts by a transmission from the incoming
lead $i$ to the bonds $(i,r)$ with transmission amplitudes $\tau^{(i)}_r$.
Multiple scattering  is induced by $(I-\tilde S(k;A))^{-1}= \sum_{n=0}^{\infty}
\tilde S ^n(k;A) $: The wave gains a phase ${\rm e}^{i(k+A_{b})L_{b}}$ for each
bond it traverses, and a scattering amplitude $\tilde \sigma^{(i)}_{r,s}$ at
each vertex, until it is transmitted from the bond $(s,j)$  to the lead $j$
with an amplitude $\tau _s^{(j)}$. Explicitly,
\begin{equation}
S^{(V)}_{i,j}  = \delta_{i,j} \rho^{(i)} +  \sum_{p \in {\cal
P}_{i\rightarrow j}}
{\cal B}_{p} {\rm e}^{i (k l_p +  b_p)}
\label {sexplicit}
\end{equation}
where ${\cal P}_{i\rightarrow j}$ is the set of the trajectories on
$\tilde {\cal G}$ which lead from $i$ to $j$.  ${\cal B}_{p}$ is the amplitude
corresponding to a path $p$  whose length and directed length are
$l_p=\sum_{b\in p}L_b$ and $b_p=\sum_{b\in p}L_bA_b$  respectively. The
scattering
amplitude $S^{(V)}_{i,j}$ is  a sum of a large number of partial
amplitudes, whose complex  interference brings about the typical irregular
fluctuations of $|S^{(V)}_{i,j}|^2$  as a function of $k$.

The resonances are the (complex) zeros of
\begin{equation}
Z_{\tilde {\cal G}}(k) = \det \left(I -\tilde S(k;A)\right) \ .
\label {resonancecond}
\end{equation}
The eigenvalues of $\tilde S$ are in the unit circle, and therefore the
resonances appear in the lower half of the complex $k$ plane.
Denoting the eigenvalues of $ S^{(V)}(k)$ by ${\rm e}^{i\theta_r(k)}$,
 $\det S^{(V)}(k) \equiv \exp [i\Theta(k)]=\exp [i
\sum_{r=1}^{V}
\theta_r(k)]$ is derived from (\ref{scatmat})  by standard manipulations\cite
{MW69}, giving
\begin{equation}
\Theta(k)- \Theta(0) = -2 {\cal I}m \log \det (I- \tilde S(k;A)) + {\cal L}k
\, .
\label{tra}
\end{equation}
${\cal L} =2\sum_{b=1}^B L_b $ is twice the total length of the bonds of
$\tilde {\cal G}$.  The resonance density $d_R(k)$ (which is proportional
to the
Wigner delay  time) \cite {S89,Krein} is given by $d_R(k) \equiv {1\over
2\pi}{{\rm d}
\Theta(k)\over {\rm d} k}$. It assigns to each resonance a normalized
Lorentzian
centered at $\Re e( k_n ) $ with a width 2 $\Im m (k_n)$. Hence,
\begin{equation}
\label{resdens}
d_R(k)  = {1\over 2 \pi}{\cal L}+{1\over \pi}{\cal
R}e\sum_{n=1}^{\infty}\sum_{p\in {\cal P}_{n}} n_p l_p r  {\tilde {\cal A}_p}^r
e^{i(l_p k+b_p)r}
\end{equation}
where the sum is over the set ${\cal P}_n$ of primitive periodic orbits whose
period $n_p$ is a divisor of $n$, with $r=n/n_p$. $l_p$ and $b_p$ are the
length
and the directed length, respectively, and the amplitudes ${\tilde {\cal A}}_p$
are the products of the bond-bond scattering amplitudes $\tilde
\sigma^i_{b,b'}$
along the primitive loops.  The mean resonance spacing is given by $\Delta=
2\pi/{\cal L}$.  (\ref {resdens}) is an exact trace formula for the
resonance density.
\begin{figure}
\hspace*{-1cm} \epsfig{figure=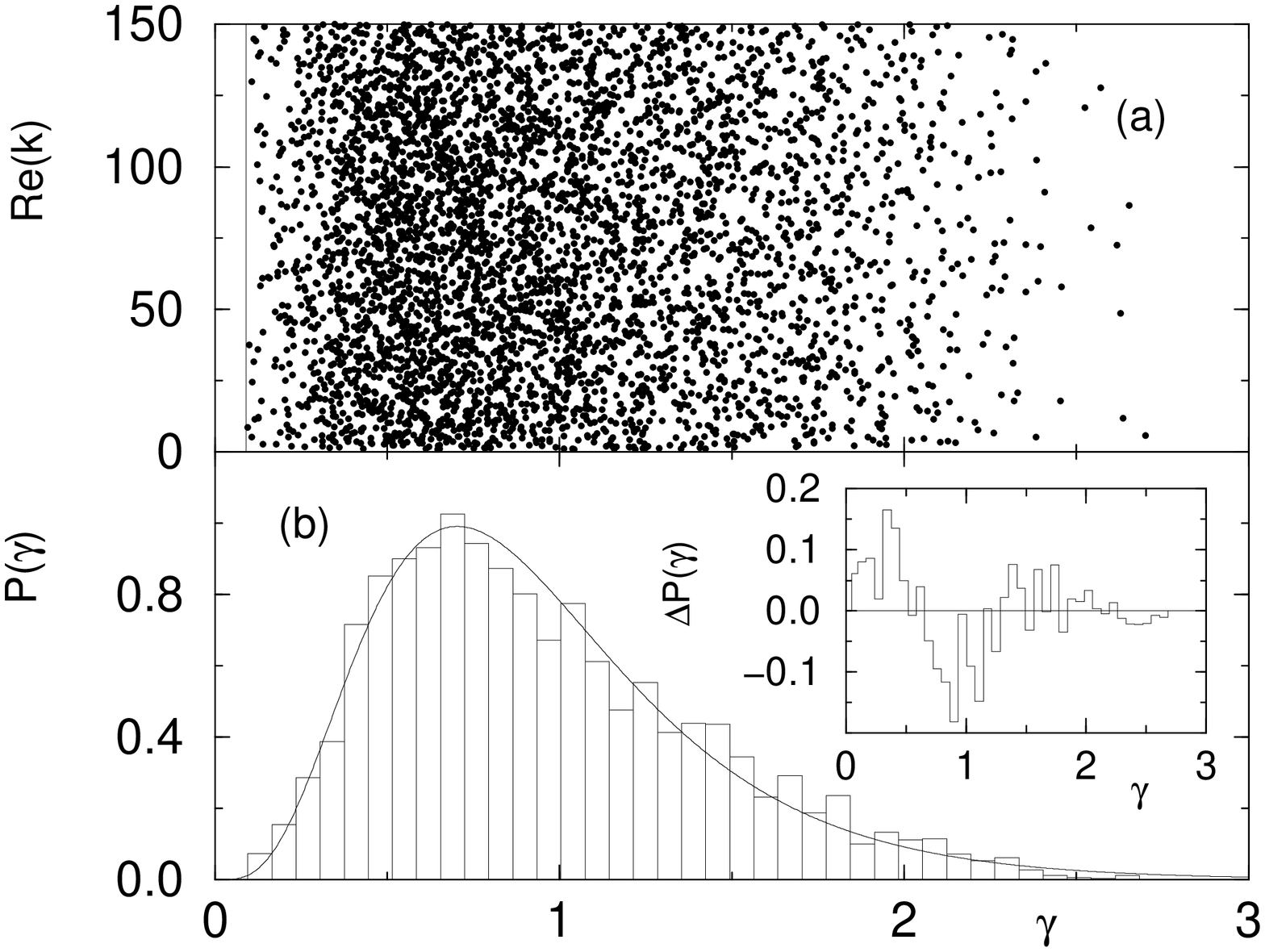,height=9cm,width=6cm,angle=270}
{\footnotesize {\bf FIG. 1.} (a) The $5000$ resonances of a single realization
of a pentagon with $A\neq 0$. The solid line marks the position of the gap
$\gamma_{gap}$. (b) The distribution ${\cal P}(\gamma)$. The solid line is the
RMT prediction \cite{FS97}. The difference ${\cal P}(\gamma)-
{\cal P}_{CUE}(\gamma)$ is shown in the inset.}
\end{figure}

The  classical dynamics  associated with  $\tilde {\cal G}$  can be easily
defined on the bonds, but not on the vertices which are singular points.
However, a Liouville description is constructed (see \cite {KS97}) by
considering the evolution of a phase-space density over the  $2B$ dimensional
space of directed bonds.  The corresponding evolution operator consists of  the
transition probabilities $\tilde U_{b,b'}$ between connected  bonds $b,b'$,
taken from the corresponding quantum evolution operator, $\tilde U_{b,b'}=
|\tilde R_{b,b'} |^2$. Due to scattering to the leads $\sum_{b'}{\tilde
U}_{bb'}<1$,
and the phase-space measure is not preserved, but rather, decays in time. Let
${\tilde p}_b(n)$ denote the probability to occupy the bond $b$ at the
(topological)
time $n$. The probability to remain on $\tilde {\cal G}$ is
\begin{equation}
{\tilde P}(n) \equiv \sum_{b=1}^{2B}{\tilde p}_b(n) = \sum_{b,b'}{\tilde
U}_{bb'}{\tilde p}_{b'}(n-1)
\simeq  e^{-\Gamma_{cl}n}{\tilde P(0)}
\end{equation}
where $\exp (-\Gamma_{cl})$ is the largest eigenvalue of the ``leaky" evolution
operator ${\tilde U}_{bb'}$. For the $v$-regular graph (\ref {Neumann}), the
probability to leak to the lead per time step is $\tau^2 $, hence,
$\Gamma_{cl} \approx (2/(1+v))^2$. The set of
trapped trajectories whose occupancy decays exponentially in time is the
analogue of the strange repeller  in generic Hamiltonian systems displaying
``chaotic scattering".

The formalism above can be easily modified for graphs where not all the
vertices
are attached to leads. If $l$,  is not attached, one has to set
$\rho^{(l)}=1, \tau^{(l)}_j =0$  in the definition of $\Sigma^{(l)}$. The
dimension of $S_{(V)}$  is  changed accordingly.

For generic
graphs, the eigenvalues of the $\tilde S$ matrix are strictly inside the unit
circle so the resonance widths $\Gamma_n \equiv -2 {\cal I}m (k_n)$, are
excluded from
the domain  $\Gamma_n \ge \Gamma_{gap}=-2 \log(|\lambda_{max}|)/L_{max}$,
where $\lambda_{max}$ is the largest eigenvalue of $\tilde S(0;A)$ and
$L_{max}$
is the longest bond. The existence of a gap - typical for chaotic
scattering \cite{molecular} from  sufficiently open scatterers- is apparent in
Fig.~1a. The widths are scaled by the
mean spacing $\Delta$ between resonances i.e. $\gamma_n\equiv {\Gamma_n \over
\Delta}$ so that $\langle \gamma\rangle_k$ determines whether the resonances
are
overlapping $(\langle \gamma\rangle_k >1)$ or isolated $(\langle
\gamma\rangle_k <1)$
( $\langle \cdot \rangle _k$ denotes spectral averaging).
The distribution of $\{\gamma_n\}$'s is shown in Fig.~1b together with the
predictions of RMT for the CUE ensemble \cite{FS97}. In spite of the general
good agreement, it deviates systematically from the numerical result (see
inset), and it does not reproduce the sharp gap in the width spectrum.
The cross section fluctuations depend crucially on $\langle \gamma\rangle_k$,
which can be approximated by the classical decay constant $\langle
\gamma\rangle_k=\gamma_{cl}$ (see Fig.~3b). For the  $v$ regular graphs
discussed
above,  $ \gamma_{cl} \approx {4\over 2\pi} {v\over 1+v} {V\over 1+v}.$
Changing $v$
and $V$ we can control the degree of overlap allowing to test various
phenomena as will be described below.
\begin{figure}
\hspace*{-1cm}\epsfig{figure=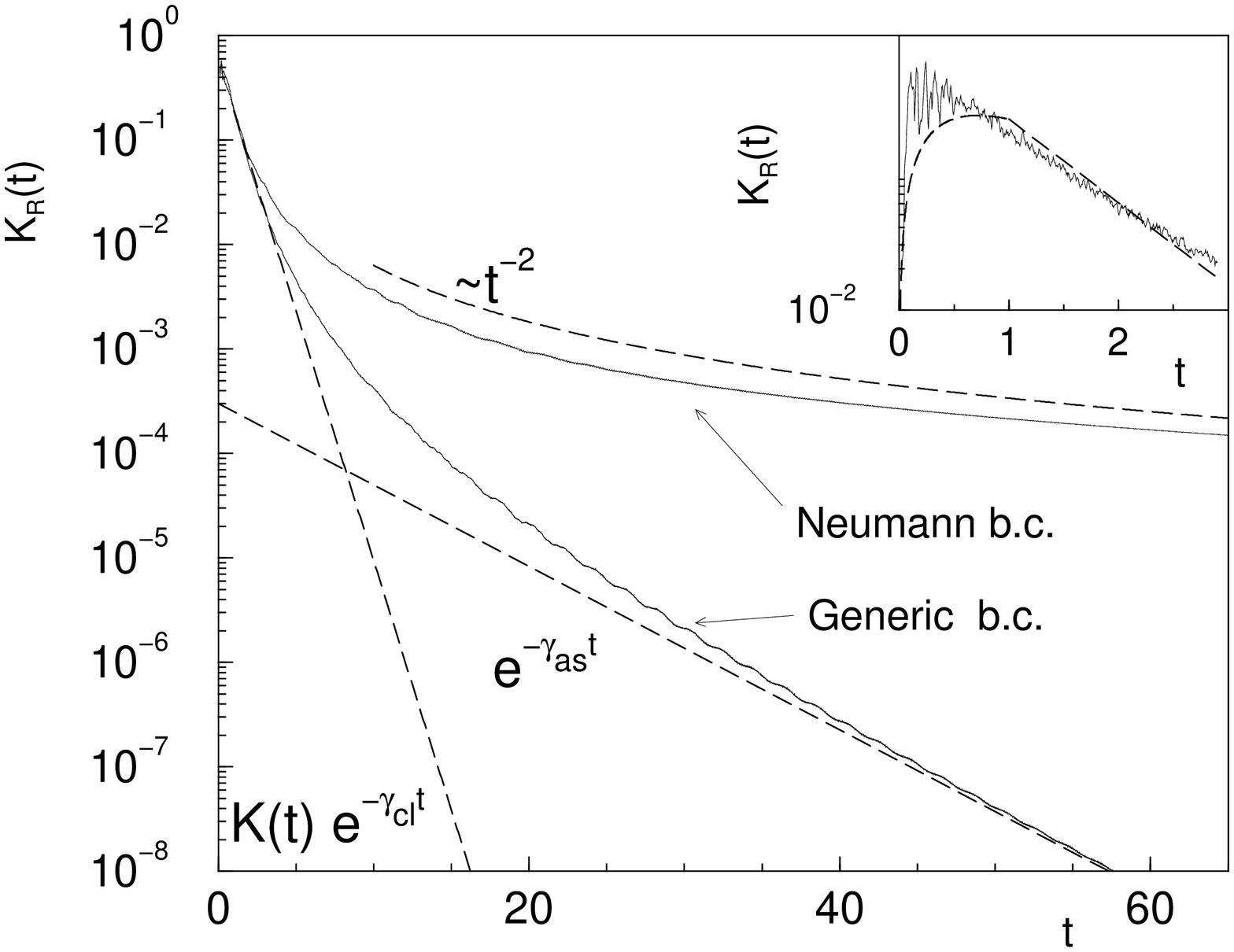,height=9cm,width=6cm,angle=270}
{\footnotesize {\bf FIG. 2.}
The form factor $K_R(t)$ for a pentagon with generic and Neumann boundary
conditions with the same mean resonance width $\left \langle
\gamma \right \rangle _k$ and $A\neq 0$. The data were averaged over $5000$
spectral intervals and smoothed on small $t$ intervals. In the inset we show
$K_R(t)$ for small times. Solid line correspond to the numerical data for
the pentagon with generic boundary conditions while dashed line is the
approximant $K_R(t)\approx K(t) exp(-\gamma_{cl} t)$. }
\end{figure}

We shall use (\ref {resdens}) to study the resonance correlation function
in terms of its form-factor
\begin{equation}
\label{rescor} K_R(t)\equiv  \int {\rm d}\chi \ {\rm e}^{i 2\pi
\chi {\cal L} t} \langle \tilde d_R  (k+ {\chi \over 2})\ \tilde d_R(k-
{\chi\over 2})\rangle_k
\end{equation}
where $\tilde d_R(k)$ is the oscillatory part of $d_R(k)$. Substituting
(\ref {resdens}), we find that the value of $K_R(t)$ equals the squared
sum of amplitudes $\tilde {\cal A}_p$ of the periodic orbits of length
$r l_p=t {\cal L}$. A similar sum contributes to the spectral from factor
of the {\it compact} graph ${\cal G}$ \cite {KS97}. The corresponding
amplitudes are different due to the fact that $\tilde {\cal A}_p$ includes
also the information about the escape of flux to the leads. Assuming that
all periodic orbits decay at the same rate, one would expect $K_R(t)
\approx K(t) e^{-\gamma_{cl}t}$, where $K(t)$ is the form factor
for the compact system \cite{Eckhardt99}. This simple approximation
is checked in the inset of Fig.~2 (see dashed line) and it is shown to
reproduce the numerical data rather well in the domain $t \le  5$. The
asymptotic decay is
dominated by the resonances which are nearest to the gap, and it cannot be
captured by the crude argument presented above. For generic graph, $K_R$ decays
exponentially but with a rate given by $\gamma_{as}=\gamma_{gap}$ (the
best fit, indicated in Fig.~2 by the dashed line, give $\gamma_{as}$ that
deviates by $30\%$ from $\gamma_{gap}$). For the graph with
Neumann boundary conditions, $\gamma_{gap}=0$ and one expects an asymptotic
power-law decay. (The corresponding dashed line in Fig.~2 shows $t^{-2}$).

Another signature of overlapping resonances are the Ericson
fluctuations observed in the $k$ dependence of the scattering cross-sections.
They are one of the prominent features which characterize generic chaotic
scattering, in the semiclassical limit.   A convenient measure for Ericson
fluctuations is the autocorrelation function
\begin{eqnarray}
\label{ericcor}
C(\chi; \nu) =  {1\over\Delta j} \sum _{ j=j_{min}}^{j_{max}}\langle  \
S^{(V)}_{j,j+\nu}(k+{\chi\over 2}) \ S^{(V)\star}_{j,j+\nu}
(k-{\chi \over 2}) \ \rangle_{k}
\end{eqnarray}
where $\Delta j = j_{max}-j_{min}+1$.
 Substituting  (\ref{sexplicit}) in
 (\ref{ericcor}) we
split the sum over trajectories into two distinct parts: the contributions
of short trajectories are computed explicitly by following the multiple
scattering expansion up to trajectories of length $l_{max}$. The contribution
of longer orbits are approximated by using the diagonal approximation, which
results in a Lorentzian with a width $\gamma_{Er}$, expected to be well
approximated by $\gamma_{cl}$. Including explicitly up to
$n=3$ scatterings we get,
\begin{eqnarray}
\label{ericscl}
&C(\chi;\nu ) &\approx G e^{il_{max}\chi} \frac {
\gamma_{Er}}{\gamma_{Er}-i\chi}
+\frac1{\Delta j}\sum_{j=j_{min}}^{j_{max}}\large[ \tau^4 e^{i\chi L_{j,j+\nu}}
\nonumber \\
&+&\tau^4 \rho^4 e^{3i\chi L_{j,j+\nu} }
+ \tau^6 \sum_{m\neq j,j+\nu} e^{i\chi(L_{j,m}+L_{m,j+\nu})} \large]
\end{eqnarray}
where the constant $G$ is determined by the normalization condition
$C(\chi=0;\nu)=1$.  The interplay between the contributions of long and short
periodic orbits is shown in Fig.~3a. For overlapping
resonances, the autocorrelation function is well reproduced by the Lorentzian
expected from the standard theory of Ericson fluctuations. The other case
corresponds to isolated resonances where the contributions of short paths are
clearly seen. From
each of the  various statistical measures of the resonance density and the
cross sections fluctuations discussed above, we  extracted  the effective
average $\gamma$, which would fit best the numerical
data.
In Fig.~3b we compare these numerical values, with the
classical expectation, and the predictions of RMT \cite{FS97}.
The results justify the use of the classical estimate for the
computation of these quantities especially in the limit
$V\rightarrow \infty$ for fixed $v/V$ (which is the analogue of
the semiclassical limit). In this limit,the RMT and
and the classical estimate coincide.
\begin{figure}
\hspace*{-1cm}\epsfig{figure=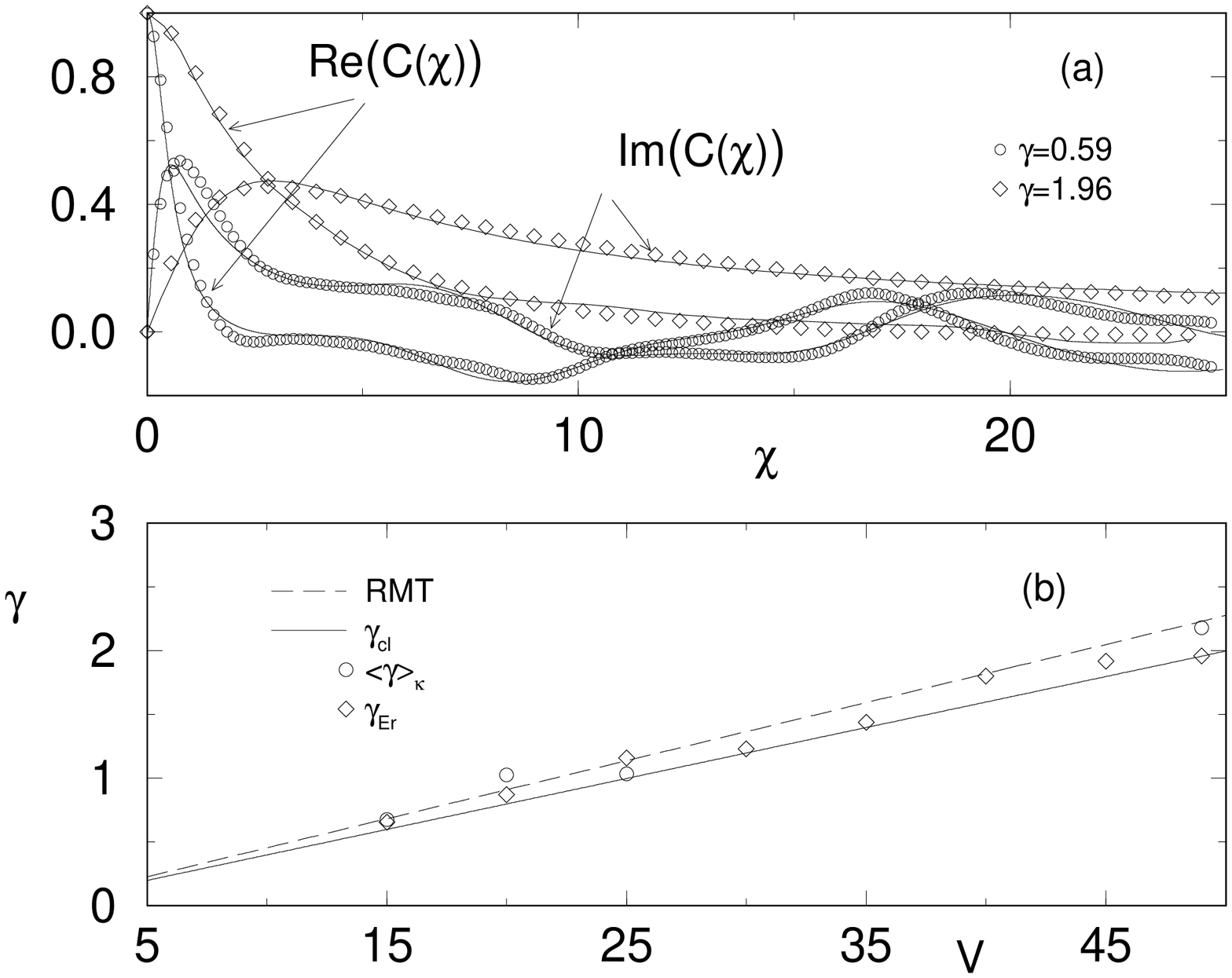,height=9cm,width=6cm,angle=270}
{\footnotesize {\bf FIG. 3.}
Regular graphs with ``Neumann" boundary conditions:
(a) The real and the imaginary part of $C(\chi,\nu=1)$ for isolated ($\circ$)
and
overlapping ($\Diamond$) resonances.  The solid lines correspond to
the theoretical expression (\ref{ericscl});
(b) Mean resonance width $\langle\gamma\rangle_k$, autocorrelation width
$\gamma_{Er}$, the classical expectation $\gamma_{cl}$ and the RMT prediction
\cite{FS97}, vs. $V$ for constant valency $v=14$.
}
\end{figure}

\begin{figure}
\hspace*{-1cm}\epsfig{figure=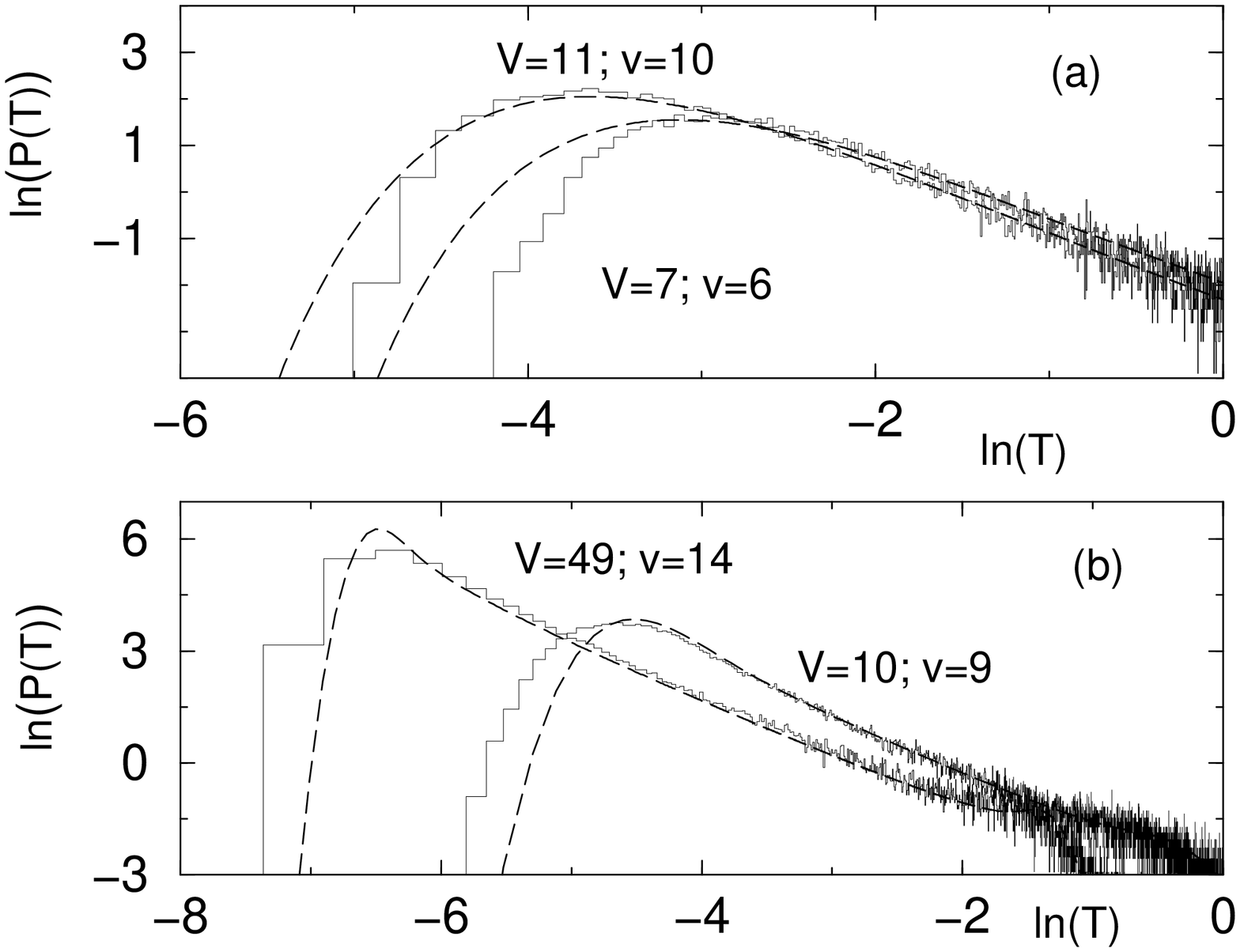,height=9cm,width=6cm,angle=270}
{\footnotesize {\bf FIG. 4.}
The distribution of the scaled partial delay times  $T$ for various graphs with
``Neumann" boundary conditions. The dashed lines correspond to the RMT
expectation
\cite{FS97}. (a) One channel and
$A=0$; (b) $V$ channels
and $A\neq 0$. }
\end{figure}
To investigate further the statistical properties of the $S^{(V)}$ matrix, we
study the distribution of scaled partial Wigner delay times $T = {\Delta \over
2\pi}{\partial \theta_r(k)\over \partial k }$. The resulting distribution for
various graphs with $A=0$ and $A\neq 0$ are shown in Figs.~4a,b respectively,
together with the predictions of RMT \cite{FS97}. An overall agreement is
evident. Deviations appear at the short time regime (i.e. short orbits),
during which the ``chaotic" component due to multiple scattering is not
yet fully developed \cite {RBUS90}.

To summarize, we presented analytical and numerical results, on the basis of
which, we propose quantum graphs as a model for the study of quantum chaotic
scattering. Their simplicity enable us to get new understanding on the subject.
Because of lack of space we defer the discussion of other results and further
comparisons with RMT to a later publication
\cite{KS00}.

This work was supported by the Minerva center for Nonlinear
Physics, and an Israel Science Foundation Grant. TK acknowledges a postdoctoral
fellowship from the Feinberg School, The Weizmann Institute of Science.

\end{multicols}
\end{document}